\documentclass[12pt]{article}
\usepackage{amssymb}
\usepackage{graphicx}
\usepackage{amsmath}

\input{tcilatex}
\input{tcilatex}
\begin{document}

\begin{center}
{\LARGE Phaseless inverse scattering problems in 3-d}

\bigskip by

Michael Victor Klibanov

University of North Carolina at Charlotte,

Charlotte, NC, U.S.A.
\end{center}

\section{Introduction}

\label{sec:1}

Consider the Schr\"{o}dinger equation in $\mathbb{R}^{3}$ with the compactly
supported potential $q\left( x\right) ,x\in \mathbb{R}^{3}$. The problem of
the reconstruction of the function $q\left( x\right) $ from measurements of
the solution of that equation on a certain set is called \textquotedblleft
inverse scattering problem". In this paper we prove uniqueness theorems for
some 3-d inverse scattering problems in the case when only the modulus of
the complex valued wave field is measured, while the phase is unknown.\ This
is the phaseless case. In the past, phaseless inverse scattering problems
were studied only in the 1-d case (section 1.2). As to the 3-d inverse
scattering problems in the frequency domain, it was assumed in all studies
so far that both the modulus and the phase of the complex valued wave field
are measured, see, e.g. \cite{Ber} for uniqueness results in the case of a
piecewise analytic potential and \cite{Nov1,Nov2}\ for global uniqueness
results and reconstruction methods.

Below $C^{s+\alpha }$ are H\"{o}lder spaces, where $s\geq 0$ is an integer
and $\alpha \in \left( 0,1\right) .$ Let $\Omega ,G\subset \mathbb{R}^{3}$
be two bounded domains, $\Omega \subset G$. For an arbitrary point $y\in 
\mathbb{R}^{3}$ and for an arbitrary number $\omega \in \left( 0,1\right) $
denote $B_{\omega }\left( y\right) =\left\{ x:\left\vert x-y\right\vert
<\omega \right\} $ and $P_{\omega }\left( y\right) =\mathbb{R}^{3}\diagdown
B_{\omega }\left( y\right) .$ For any two sets $M,N\subset \mathbb{R}^{3}$
let $dist\left( M,N\right) $ be the Hausdorff distance between them. Let $%
G_{1}\subset \mathbb{R}^{3}$ be a convex bounded domain with its boundary $%
S\in C^{1}$.\ Let $\varepsilon \in \left( 0,1\right) $ be a number. We
assume that $\Omega \subset G_{1}\subset G,$ $dist\left( S,\partial G\right)
>2\varepsilon $ and $dist\left( S,\partial \Omega \right) >2\varepsilon $.
Hence, 
\begin{eqnarray}
dist\left( \partial B_{\varepsilon }\left( y\right) ,\partial \Omega \right)
&>&\varepsilon ,\forall y\in S,  \label{1} \\
dist\left( \partial B_{\varepsilon }\left( y\right) ,\partial G\right)
&>&\varepsilon ,\forall y\in S.  \label{2}
\end{eqnarray}
Below either $m=2$ or $m=4,$ and we will specify this later. We impose the
following conditions on the potential $q\left( x\right) $ 
\begin{eqnarray}
q\left( x\right) &\in &C^{m}\left( \mathbb{R}^{3}\right) ,q\left( x\right) =0%
\text{ for }x\in \mathbb{R}^{3}\diagdown G,  \label{1.1} \\
q\left( x\right) &\geq &0.  \label{1.2}
\end{eqnarray}

As a rule, the minimal smoothness of unknown coefficients is not the first
priority of proofs of uniqueness theorems of multidimensional coefficient
inverse problems, see, e.g. \cite{Nov1,Nov2} and Theorem 4.1 in \cite{Rom2}.
Since our proofs require either $C^{2}$ or $C^{4}$ smoothness of solutions
of Cauchy problems for some hyperbolic equations, we are not concerned below
with minimal smoothness assumptions. In particular, the reason of imposing $%
C^{4}$ (rather than $C^{2}$) smoothness conditions in Theorems 3 and 4 is
rooted in smoothness requirements of uniqueness theorems of \cite%
{BK,BukhKlib,Bukh2,Klib5,KT,Kl16}, which we use here.

\subsection{One of main results}

\label{sec:1.1}

We now formulate one of our four main theorems. Three other theorems are
formulated in section 2. Let $x_{0}=\left( x_{0,1},x_{0,2},x_{0,3}\right) $
be the source position. Consider the following problem 
\begin{equation}
\Delta _{x}u+k^{2}u-q\left( x\right) u=-\delta \left( x-x_{0}\right) ,x\in 
\mathbb{R}^{3},  \label{1.4}
\end{equation}
\begin{equation}
u\left( x,x_{0},k\right) =O\left( \frac{1}{\left| x-x_{0}\right| }\right)
,\left| x\right| \rightarrow \infty ,  \label{1.50}
\end{equation}
\begin{equation}
\sum\limits_{j=1}^{3}\frac{x_{j}-x_{j,0}}{\left| x-x_{0}\right| }\partial
_{x_{j}}u\left( x,x_{0},k\right) -iku\left( x,x_{0},k\right) =o\left( \frac{1%
}{\left| x-x_{0}\right| }\right) ,\left| x\right| \rightarrow \infty .
\label{1.5}
\end{equation}
Here the radiation conditions (\ref{1.50}), (\ref{1.5}) are valid for every
fixed source position $x_{0}.$ To establish existence and uniqueness of the
solution of the problem (\ref{1.4})-(\ref{1.5}), we refer to Theorem 6 of
Chapter 9 of the book \cite{V} as well as to Theorem 3.3 of the paper \cite%
{V1}. As to the smoothness of the solution of the problem (\ref{1.4})-(\ref%
{1.5}), we refer to Theorem 6.17 of the book \cite{GT}. \ Thus, combining
these results, we obtain that for each pair $\left( k,x_{0}\right) \in 
\mathbb{R\times R}^{3}$ there exists unique solution $u\left(
x,x_{0},k\right) $ of the problem (\ref{1.4}), (\ref{1.50}), (\ref{1.5})
such that 
\begin{equation}
u\left( x,x_{0},k\right) =u_{0}\left( x,x_{0},k\right) +u_{s}\left(
x,x_{0},k\right) ,  \label{1.6}
\end{equation}
\begin{equation}
u_{0}=\frac{\exp \left( ik\left| x-x_{0}\right| \right) }{4\pi \left|
x-x_{0}\right| },\text{ }u_{s}\in C^{m+1+\alpha }\left( P_{\omega }\left(
x_{0}\right) \right) ,\forall \alpha ,\omega \in \left( 0,1\right) .
\label{1.7}
\end{equation}
In (\ref{1.6}), (\ref{1.7}) $u_{0}\left( x,x_{0},k\right) $ is the incident
spherical wave and $u_{s}\left( x,x_{0},k\right) $ is the scattered wave.

\textbf{Inverse Problem 1 (IP1)}. Let $m=2$ in (\ref{1.1}). Suppose that the
function $q\left( x\right) $ satisfying (\ref{1.1}), (\ref{1.2}) is unknown
for $x\in \Omega $ and known for $x\in \mathbb{R}^{3}\diagdown \Omega .$
Also, assume that the following function $f_{1}\left( x,x_{0},k\right) $ is
known 
\begin{equation*}
f_{1}\left( x,x_{0},k\right) =\left\vert u\left( x,x_{0},k\right)
\right\vert ^{2},\forall x_{0}\in S,\forall x\in B_{\varepsilon }\left(
x_{0}\right) ,x\neq x_{0},\forall k\in \left( a,b\right) ,
\end{equation*}
where $\left( a,b\right) \subset \mathbb{R}$ is an arbitrary interval.
Determine the function $q\left( x\right) $ for $x\in \Omega .$

Theorem 1 is one of four main results of this paper.

\textbf{Theorem 1}. \emph{Consider IP1.} \emph{Let two potentials }$%
q_{1}\left( x\right) $ \emph{and }$q_{2}\left( x\right) $\emph{\ satisfying
conditions (\ref{1.1}), (\ref{1.2}) be such that }$q_{1}\left( x\right)
=q_{2}\left( x\right) =q\left( x\right) $ \emph{for} $x\in \mathbb{R}%
^{3}\diagdown \Omega .$ \emph{Let }$u_{1}\left( x,x_{0},k\right) $ \emph{and}
$u_{2}\left( x,x_{0},k\right) $\emph{\ be corresponding solutions of the
problem (\ref{1.4})-(\ref{1.5}) satisfying conditions (\ref{1.6}), (\ref{1.7}%
). Assume that } 
\begin{equation}
\left\vert u_{1}\left( x,x_{0},k\right) \right\vert ^{2}=\left\vert
u_{2}\left( x,x_{0},k\right) \right\vert ^{2},\forall x_{0}\in S,\forall
x\in B_{\varepsilon }\left( x_{0}\right) ,x\neq x_{0},\forall k\in \left(
a,b\right) .  \label{1.19}
\end{equation}
\emph{Then }$q_{1}\left( x\right) \equiv q_{2}\left( x\right) .$

\textbf{Corollary 1}. \emph{Fix two arbitrary points }$y_{0}\in S$ \emph{and}
$y\in B_{\varepsilon }\left( y_{0}\right) $ \emph{such that }$y\neq y_{0}$%
\emph{. Suppose that all conditions of Theorem 1 are in place, except that (%
\ref{1.19}) is replaced with } 
\begin{equation}
\left\vert u_{1}\left( y,y_{0},k\right) \right\vert ^{2}=\left\vert
u_{2}\left( y,y_{0},k\right) \right\vert ^{2}\emph{,}\text{ }\forall k\in
\left( a,b\right) .  \label{1.15}
\end{equation}
\emph{\ Then }$u_{1}\left( y,y_{0},k\right) =u_{2}\left( y,y_{0},k\right) $%
\emph{\ for all }$k\in \mathbb{R}$\emph{.}

\textbf{Remark 1}. The proof of Corollary 1 can be immediately derived from
the proof of Theorem 1. Completely analogous corollaries are valid for each
of Theorems 2-4 of this paper. Their proofs can also be immediately derived
from proofs of corresponding theorems. We omit formulations of those
corollaries for brevity.

We now outline the main difficulty, which did not allow to prove uniqueness
results for phaseless 3-d inverse scattering problems so far. As an example
we consider IP1. Analogous difficulties take place for three other inverse
problems formulated in section 2. In IP1 one should work with a complex
valued function $r\left( k\right) ,k\in \mathbb{R}$ such that its modulus $%
\left\vert r\left( k\right) \right\vert $ is known for all $k\in \left(
a,b\right) $. The function $r\left( k\right) $ admits the analytic
continuation from the real line $\mathbb{R}$ in the half-plane $\left\{ k\in 
\mathbb{C}:\func{Im}k>-\gamma \right\} $ for a certain number $\gamma >0.$
Since $\left\vert r\left( k\right) \right\vert ^{2}=r\left( k\right) 
\overline{r}\left( k\right) ,$ then the function $\left\vert r\left(
k\right) \right\vert ^{2}$ is analytic for $k\in \mathbb{R}$ as the function
of the real variable $k$. Here $\overline{r}\left( k\right) $ is the complex
conjugate of $r\left( k\right) .$ Hence, the modulus $\left\vert r\left(
k\right) \right\vert $ is known for all $k\in \mathbb{R}.$ Denote $\mathbb{C}%
_{+}=\left\{ k\in \mathbb{C}:\func{Im}k\geq 0\right\} .$ Proposition 4.2 of 
\cite{Kl11} implies that if $r\left( k\right) $ would not have zeros in $%
\mathbb{C}_{+},$ then this function would be uniquely reconstructed for all $%
k\in \mathbb{R}$ from the values of $\left\vert r\left( k\right) \right\vert 
$ for $k\in \mathbb{R}$, also see Lemma 4 in subsection 3.2. However, the 
\emph{main difficulty} is to properly account for zeros of $r\left( k\right) 
$ in the upper half-plane $\mathbb{C}_{+}\diagdown \mathbb{R}.$ Indeed, let $%
z_{1},...,z_{n}\in \mathbb{C}_{+}\diagdown \mathbb{R}$ be some of such zeros
of $r\left( k\right) .$ Consider the function $\widehat{r}\left( k\right) $
defined as 
\begin{equation*}
\widehat{r}\left( k\right) =\left( \prod\limits_{j=1}^{n}\frac{k-\overline{z}%
_{j}}{k-z_{j}}\right) r\left( k\right) .
\end{equation*}
Hence, $\left\vert \widehat{r}\left( k\right) \right\vert =\left\vert
r\left( k\right) \right\vert ,\forall k\in \mathbb{R}$. Furthermore, the
function $\widehat{r}\left( k\right) $ is analytic in the half-plane

$\left\{ k\in \mathbb{C}:\func{Im}k>-\gamma \right\} .$ Therefore, in order
to prove uniqueness, one needs to figure out how to combine the knowledge of 
$\left| r\left( k\right) \right| $ for $k\in \mathbb{R}$ with a linkage
between the function $r\left( k\right) $ and the originating differential
operator.

This difficulty was handled in \cite{Kl9} in 1-d, using the fact that the
function $r\left( k\right) $ depends only on one variable $k$ in this case.
Unlike \cite{Kl9}, the function $r\left( x,x_{0},k\right) =u\left(
x,x_{0},k\right) $ depends on $x,x_{0},k$ in the 3-d case. Hence, the above
zeros depend now on both $x$ and $x_{0},$ i.e. $z_{j}=z_{j}\left(
x,x_{0}\right) .$ Thus, compared with the 1-d problem, the main difficulty
of the 3-d case is that it is necessary to figure out how to take into
account the dependence of zeros $z_{j}\left( x,x_{0}\right) $ from $x$ and $%
x_{0}$. To do this, we essentially use here properties of the solution of
the Cauchy problem for an associated hyperbolic PDE.

\subsection{Published results}

\label{sec:1.2}

The phaseless inverse scattering problem is of central importance in some
applications, where only the amplitude of the scattered signal can be
measured. An example is neutron specular reflection, see, e.g. \cite{BM}.
Uniqueness of the phaseless inverse scattering problem in the 1-d case was
first proved in \cite{Kl9}. Next, the result of \cite{Kl9} was extended to
the discontinuous impedance case in \cite{NHS}. Also, see \cite{AS} for a
relevant result. A survey can be found in \cite{Kl11}. Uniqueness theorem
for a 1-d phaseless inverse problem arising in crystallography was proven in 
\cite{K}. This problem is essentially different from the one considered in 
\cite{Kl9}.

Inverse problems without the phase information are well known in optics,
since it is often impossible to measure the phase of the optical signal,
unlike its amplitude. In optics, such a problem is usually formulated as the
problem about the recovery of a compactly supported complex valued function
from the modulus of its Fourier transform.~The latter is called the
\textquotedblleft phase retrieval problem" \cite{Fienup1}. This problem
arises in x-ray crystallography \cite{Ladd}, astronomical imaging \cite%
{Fienup2} and other subfields of optics \cite{Fienup1}. Some numerical
methods for this problem can be found in, e.g. \cite{Dob,Fienup1,Fienup2,H,S}%
. Recently regularization algorithms were developed in the 1-d case for a
similar, the so-called \textquotedblleft autocorrelation problem" \cite%
{DaiLamm,Hoffman}. Uniqueness theorems for the phase retrieval problem can
be found in \cite{Kl8,Kl12}.

In section 2 we formulate three more phaseless inverse scattering problems
as well uniqueness theorems 2-4 for them. In section 3 we prove Theorem 1.
Theorem 2 is proved in section 4. Finally, Theorems 3 and 4 are proved in
section 5.

\section{Other problems and results}

\label{sec:2}

In IP1 the modulus of the total wave field $u=u_{0}+u_{s}$ is known on a
certain set. We now consider the case when the modulus of the scattered wave
is known.

\textbf{Inverse Problem 2 (IP2)}. Let $m=2$ in (\ref{1.1}). Suppose that the
function $q\left( x\right) $ satisfying (\ref{1.1}), (\ref{1.2}) is unknown
for $x\in \Omega $ and known for $x\in \mathbb{R}^{3}\diagdown \Omega .$
Also, assume that the following function $f_{2}\left( x,x_{0},k\right) $ is
known 
\begin{equation*}
f_{2}\left( x,x_{0},k\right) =\left\vert u_{s}\left( x,x_{0},k\right)
\right\vert ^{2},\forall x_{0}\in S,\forall x\in B_{\varepsilon }\left(
x_{0}\right) ,x\neq x_{0},\forall k\in \left( a,b\right) .
\end{equation*}
Determine the function $q\left( x\right) $ for $x\in \Omega .$

\textbf{Theorem 2}. \emph{Consider IP2. Assume that all conditions of
Theorem 1 hold, except that (\ref{1.19}) is replaced with} 
\begin{equation}
\left\vert u_{s,1}\left( x,x_{0},k\right) \right\vert ^{2}=\left\vert
u_{s,2}\left( x,x_{0},k\right) \right\vert ^{2},\forall x_{0}\in S,\forall
x\in B_{\varepsilon }\left( x_{0}\right) ,x\neq x_{0},\forall k\in \left(
a,b\right) ,  \label{1.20}
\end{equation}
\emph{where }$u_{s,j}=u_{j}-u_{0},j=1,2.$ \emph{In addition, assume that }$%
q\left( x\right) \neq 0$\emph{, }$\forall x\in S.$ \emph{Then }$q_{1}\left(
x\right) \equiv q_{2}\left( x\right) .$

Theorems 1 and 2 are formulated only for the over-determined data. Indeed,
in both IP1 and IP2 the number of free variables in the data exceeds the
number of free variables in the unknown coefficient. The reason of this is
that even if the phase is known, still all current uniqueness results for
3-d inverse scattering problems in the case when the $\delta -$function is
the source function are valid only if the data are over-determined ones,
see, e.g. \cite{Ber,Nov1,Nov2} for the frequency domain and \S 1 of chapter
7 of \cite{LRS} for an inverse scattering problem in the time domain.
Suppose now that the function $\delta \left( x-x_{0}\right) $ in (\ref{1.4})
is replaced with such a function $p\left( x\right) $ that $p\left( x\right)
\neq 0$ in $\overline{\Omega }.$ And consider the inverse problem of the
reconstruction of the potential $q\left( x\right) $ from values of the
function $u\left( x,k\right) $ for all $x\in S,k\in \mathbb{R}$. Then
uniqueness theorem for this problem can be proved for the non-overdetermined
case. This proof can be handled by the method, which was introduced in the
originating paper \cite{BukhKlib}. Also, see, e.g. \cite{Bukh2,Klib5,KT} and
sections 1.10, 1.11 of \cite{BK} for some follow up works of authors of \cite%
{BukhKlib} on this method; a survey can be found in \cite{Kl16}. This
technique is based on Carleman estimates.

Consider the function $\chi \left( x\right) \in C^{\infty }\left( \mathbb{R}%
^{3}\right) $ such that $\chi \left( x\right) =1$ in $G_{1}$ and $\chi
\left( x\right) =0$ for $x\notin G.$ Let $x_{0}\in S.$ For a number $\sigma
>0$ consider the function $\delta _{\sigma }\left( x-x_{0}\right) ,$%
\begin{equation*}
\delta _{\sigma }\left( x-x_{0}\right) =C\frac{\chi \left( x\right) }{\left(
2\sqrt{\pi \sigma }\right) ^{3}}\exp \left( -\frac{\left\vert
x-x_{0}\right\vert ^{2}}{4\sigma }\right) ,
\end{equation*}
where the number $C>0$ is such that 
\begin{equation*}
\int\limits_{G}\delta _{\sigma }\left( x-x_{0}\right) dx=1.
\end{equation*}
The function $\delta _{\sigma }\left( x-x_{0}\right) $ approximates the
function $\delta \left( x-x_{0}\right) $ in the distribution sense for
sufficiently small values of $\sigma $. The function $\delta _{\sigma
}\left( x-x_{0}\right) $ is acceptable in Physics as a proper replacement of 
$\delta \left( x-x_{0}\right) $, since there is no \textquotedblleft true"
delta-function in the reality. On the other hand, the above mentioned method
of \cite{BukhKlib} is applicable to the case when $\delta \left(
x-x_{0}\right) $ is replaced with $\delta _{\sigma }\left( x\right) $.
Therefore, it seems to be worthy from the Physics standpoint to consider
Inverse Problems 3,4 below.

Let in (\ref{1.1}) $m=4$. To apply results, which follow from the method of 
\cite{BukhKlib}, consider the function $g\left( x\right) $ such that 
\begin{equation}
g\in C^{4}\left( \mathbb{R}^{3}\right) ,g\left( x\right) =0\text{ in }%
\mathbb{R}^{3}\diagdown G,  \label{1.10}
\end{equation}
\begin{equation}
g\left( x\right) \neq 0\text{ in }\overline{G}_{1}.  \label{1.11}
\end{equation}
Consider the following problem 
\begin{equation}
\Delta v+k^{2}v-q\left( x\right) v=-g\left( x\right) ,x\in \mathbb{R}^{3},
\label{1.12}
\end{equation}
\begin{equation}
v\left( x,k\right) =O\left( \frac{1}{\left\vert x\right\vert }\right)
,\left\vert x\right\vert \rightarrow \infty ,  \label{1.120}
\end{equation}
\begin{equation}
\sum\limits_{j=1}^{3}\frac{x_{j}}{\left\vert x\right\vert }\partial
_{x_{j}}v\left( x,k\right) -ikv\left( x,k\right) =o\left( \frac{1}{%
\left\vert x\right\vert }\right) ,\left\vert x\right\vert \rightarrow \infty
.  \label{1.13}
\end{equation}
The same results of \cite{GT,V1,V} as ones in subsection 1.1 guarantee that
for each $k\in \mathbb{R}$ there exists unique solution $v\left( x,k\right)
\in C^{5+\alpha }\left( \mathbb{R}^{3}\right) ,\forall \alpha \in \left(
0,1\right) $ of the problem (\ref{1.12}), (\ref{1.120}), (\ref{1.13}).

\textbf{Inverse Problem 3 (IP3)}. Let $m=4$ in (\ref{1.1}). Suppose that the
function $q\left( x\right) $ satisfying conditions (\ref{1.1}), (\ref{1.2})
is unknown for $x\in \Omega $ and known for $x\in \mathbb{R}^{3}\diagdown
\Omega .$ Assume that the following function $f_{3}\left( x,k\right) $ is
known 
\begin{equation}
f_{3}\left( x,k\right) =\left\vert v\left( x,k\right) \right\vert
^{2},\forall x\in S,\forall k\in \left( a,b\right) .  \label{1.14}
\end{equation}
Determine the function $q\left( x\right) $ for $x\in \Omega .$

\textbf{Theorem 3}. \emph{Consider IP3.} \emph{Let the function }$g\left(
x\right) $\emph{\ satisfies conditions (\ref{1.10}), (\ref{1.11}). Consider
two functions }$q_{1}\left( x\right) ,q_{2}\left( x\right) $\emph{\
satisfying conditions (\ref{1.1}), (\ref{1.2}) and such that }$q_{1}\left(
x\right) =q_{2}\left( x\right) =q\left( x\right) $ \emph{for} $x\in \mathbb{R%
}^{3}\diagdown \Omega .$\emph{\ For }$j=1,2$ \emph{let }$v_{j}\left(
x,k\right) \in C^{5+\alpha }\left( \mathbb{R}^{3}\right) $\emph{\ be the
solution of the problem (\ref{1.12})-(\ref{1.13}) with }$q\left( x\right)
=q_{j}\left( x\right) $\emph{.} \emph{Assume that\ } 
\begin{equation}
\left\vert v_{1}\left( x,k\right) \right\vert ^{2}=\left\vert v_{2}\left(
x,k\right) \right\vert ^{2},\forall x\in S,\forall k\in \left( a,b\right) .
\label{1.21}
\end{equation}
\emph{Then }$q_{1}\left( x\right) \equiv q_{2}\left( x\right) .$

We now pose an analog of IP2. Let $v_{0}\left( x,k\right) $ be the solution
of the problem (\ref{1.12})-(\ref{1.13}) for the case $q\left( x\right)
\equiv 0,$%
\begin{equation*}
v_{0}\left( x,k\right) =\int\limits_{G}\frac{\exp \left( ik\left\vert x-\xi
\right\vert \right) }{4\pi \left\vert x-\xi \right\vert }g\left( \xi \right)
d\xi .
\end{equation*}
Hence, one can interpret the function $v_{0}\left( x,k\right) $ as the
solution of the problem (\ref{1.12})-(\ref{1.13}) for case of the background
medium.

\textbf{Inverse Problem 4 (IP4)}. Let $m=4$ in (\ref{1.1}). Suppose that the
function $q\left( x\right) $ satisfying (\ref{1.1}), (\ref{1.2}) is unknown
for $x\in \Omega $ and known for $x\in \mathbb{R}^{3}\diagdown \Omega .$ Let 
$v_{s}\left( x,k\right) =v\left( x,k\right) -v_{0}\left( x,k\right) .$
Assume that the following function $f_{4}\left( x,k\right) $ is known 
\begin{equation*}
f_{4}\left( x,k\right) =\left\vert v_{s}\left( x,k\right) \right\vert
^{2},\forall x\in S,\forall k\in \left( a,b\right) .
\end{equation*}
Determine the function $q\left( x\right) $ for $x\in \Omega .$

\textbf{Theorem 4}. \emph{Consider IP4. Let all conditions of Theorem 3
hold, except that (\ref{1.21}) is replaced with} 
\begin{equation}
\left\vert v_{s,1}\left( x,k\right) \right\vert ^{2}=\left\vert
v_{s,2}\left( x,k\right) \right\vert ^{2},\forall x\in S,\forall k\in \left(
a,b\right) ,  \label{1.22}
\end{equation}
\emph{where }$v_{s,j}\left( x,k\right) =v_{j}\left( x,k\right) -v_{0}\left(
x,k\right) ,j=1,2.$ \emph{Assume that }$q\left( x\right) \neq 0,\forall x\in
S.$\emph{\ Then }$q_{1}\left( x\right) \equiv q_{2}\left( x\right) .$

\section{Proof of Theorem 1}

\label{sec:3}

\subsection{Functions $U$ and $u$}

\label{sec:3.1}

Consider the solution $U\left( x,x_{0},t\right) $ of the following Cauchy
problem 
\begin{equation}
U_{tt}=\Delta _{x}U-q\left( x\right) U,\left( x,t\right) \in \mathbb{R}%
^{3}\times \left( 0,\infty \right) ,  \label{2.1}
\end{equation}
\begin{equation}
U\left( x,0\right) =0,U_{t}\left( x,0\right) =\delta \left( x-x_{0}\right) .
\label{2.2}
\end{equation}
It was shown in \S 1 of Chapter 7 of the book \cite{LRS} that the function $%
U $ has the form 
\begin{equation}
U\left( x,x_{0},t\right) =\frac{\delta \left( t-\left\vert
x-x_{0}\right\vert \right) }{4\pi \left\vert x-x_{0}\right\vert }+\widetilde{%
U}\left( x,x_{0},t\right) ,  \label{2.3}
\end{equation}
where 
\begin{equation}
\widetilde{U}\left( x,x_{0},t\right) =-\int\limits_{D\left( x,x_{0},t\right)
}\frac{q\left( \xi \right) U\left( \xi ,x_{0},t-\left\vert x-\xi \right\vert
\right) }{4\pi \left\vert x-\xi \right\vert }d\xi ,  \label{2.31}
\end{equation}
\begin{equation}
D\left( x,x_{0},t\right) =\left\{ \xi :\left\vert x-\xi \right\vert
+\left\vert x_{0}-\xi \right\vert <t\right\} ,  \label{2.32}
\end{equation}
\begin{equation}
\widetilde{U}\left( x,x_{0},t\right) =0\text{ for }t\in \left( 0,\left\vert
x-x_{0}\right\vert \right) ,  \label{2.33}
\end{equation}
\begin{equation}
\widetilde{U}\left( x,x_{0},t\right) =-\frac{1}{16\pi \left\vert
x-x_{0}\right\vert }\int\limits_{L\left( x,x_{0}\right) }q\left( \xi \right)
ds+O\left( t-\left\vert x-x_{0}\right\vert \right) ,t\rightarrow \left\vert
x-x_{0}\right\vert ^{+}.  \label{2.4}
\end{equation}
Here $L\left( x,x_{0}\right) $ is the interval of the straight line
connecting points $x,x_{0}$.

Let $\Phi \subset \mathbb{R}^{3}$ be an arbitrary bounded domain. Choose the
number $\omega _{0}=\omega _{0}\left( x_{0},\Phi \right) \in \left(
0,1\right) $ so small that $\Phi \cap P_{\omega }\left( x_{0}\right) \neq
\varnothing .$ Let $T>\max_{x\in \overline{\Phi }}\left\vert
x-x_{0}\right\vert $ be an arbitrary number. Denote 
\begin{equation*}
\Psi \left( \Phi ,x_{0},\omega ,T\right) =\left\{ \left( x,t\right) :x\in 
\overline{\Phi }\cap P_{\omega }\left( x_{0}\right) ,t\in \left[ \left\vert
x-x_{0}\right\vert ,T\right] \right\} .
\end{equation*}
It was shown in \S 3 of chapter 2 of the book \cite{Rom1} that the function $%
U$ can be represented as the following series 
\begin{equation}
U\left( x,x_{0},t\right) =\sum\limits_{n=0}^{\infty }U_{n}\left(
x,x_{0},t\right) ,  \label{2.401}
\end{equation}
where 
\begin{equation}
U_{0}\left( x,x_{0},t\right) =\frac{\delta \left( t-\left\vert
x-x_{0}\right\vert \right) }{4\pi \left\vert x-x_{0}\right\vert },
\label{2.402}
\end{equation}
\begin{equation}
U_{n}\left( x,x_{0},t\right) =-\int\limits_{D\left( x,x_{0},t\right) }\frac{%
q\left( \xi \right) U_{n-1}\left( \xi ,x_{0},t-\left\vert x-\xi \right\vert
\right) }{4\pi \left\vert x-\xi \right\vert }d\xi ,n\geq 1.  \label{2.403}
\end{equation}
Convergence estimates of \S 3 of chapter 2 of \cite{Rom1} imply that series (%
\ref{2.401}) converges in the norm of the space $C^{2}\left( \Psi \left(
\Phi ,x_{0},\omega ,T\right) \right) $. Hence, for any fixed source position 
$x_{0}\in \mathbb{R}^{3}$ and for $\left\vert \beta \right\vert =0,1,2$ 
\begin{equation}
D_{x,t}^{\beta }\widetilde{U}\left( x,x_{0},t\right) \in C\left( \Psi \left(
\Phi ,x_{0},\omega ,T\right) \right) ,\forall \omega \in \left( 0,\omega
_{0}\left( x_{0},\Phi \right) \right) ,\forall T>\max_{\overline{\Phi }%
}\left\vert x-x_{0}\right\vert .  \label{2.41}
\end{equation}

We now refer to some results about the asymptotic behavior of solutions of
hyperbolic equations as $t\rightarrow \infty .$ More precisely, we refer to
Lemma 6 of Chapter 10 of the book \cite{V} as well as to Remark 3 after that
lemma. It follows from these results that there exist numbers $%
C_{1}=C_{1}\left( q,\Phi ,x_{0},\omega \right) >0,c_{1}=c_{1}\left( q,\Phi
,x_{0},\omega \right) >0$ depending only on the function $q$, the domain $%
\Phi $, the source position $x_{0}$ and the number $\omega \in \left(
0,\omega _{0}\left( x_{0},\Phi \right) \right) $ such that 
\begin{equation}
\left\vert D_{x,t}^{\beta }\widetilde{U}\left( x,x_{0},t\right) \right\vert
\leq C_{1}e^{-c_{1}t}\text{ in }\left\{ \left( x,t\right) :x\in \Phi \cap
P_{\omega }\left( x_{0}\right) ,t\geq \left\vert x-x_{0}\right\vert \right\}
,\left\vert \beta \right\vert =0,1,2.  \label{2.6}
\end{equation}

It follows from (\ref{2.3}), (\ref{2.41}) and (\ref{2.6}) that one can apply
the operator $\mathcal{F}$ of the Fourier transform with respect to $t$ to
functions $D_{x,t}^{\beta }\widetilde{U}\left( x,x_{0},t\right) ,\left\vert
\beta \right\vert =0,1,2.$ Let 
\begin{equation*}
\mathcal{F}\left( U\right) \left( x,x_{0},k\right) =\int\limits_{0}^{\infty
}U\left( x,x_{0},t\right) e^{ikt}dt,\text{ }\forall x,x_{0}\in \mathbb{R}%
^{3},x\neq x_{0},\forall k\in \mathbb{R}.
\end{equation*}
Using again the same results of references \cite{GT,V1,V} as ones cited in
section 1.1, we obtain that 
\begin{equation}
u\left( x,x_{0},k\right) =\mathcal{F}\left( U\right) \left( x,x_{0},k\right)
,\forall x,x_{0}\in \mathbb{R}^{3},x\neq x_{0},\forall k\in \mathbb{R}.
\label{2.7}
\end{equation}
In particular, (\ref{2.6}) and (\ref{2.7}) imply that for each pair $%
x,x_{0}\in \mathbb{R}^{3}$ such that $x\neq x_{0}$ there exists a number $%
\gamma =\gamma \left( x,x_{0},q\right) >0$ such that the function $u\left(
x,x_{0},k\right) $ admits the analytic continuation with respect to $k$ from
the real line $\mathbb{R}$ in the half-plane $\left\{ k\in \mathbb{C}:\func{%
Im}k>-\gamma \right\} $.

Suppose that $G\cap P_{\omega }\left( x_{0}\right) \neq \varnothing .$ Then,
using (\ref{2.3})-(\ref{2.4}), (\ref{2.6}) and (\ref{2.7}), we obtain the
following asymptotic behavior of functions $u\left( x,x_{0},k\right) $ and $%
u_{s}\left( x,x_{0},k\right) $ for every fixed source position $x_{0}\in 
\mathbb{R}^{3}$ and for every fixed value of $\omega \in \left( 0,\omega
_{0}\left( x_{0},\Phi \right) \right) $ 
\begin{equation}
u\left( x,x_{0},k\right) =\frac{\exp \left( ik\left\vert x-x_{0}\right\vert
\right) }{4\pi \left\vert x-x_{0}\right\vert }\left[ 1+O\left( \frac{1}{k}%
\right) \right] ,\left\vert k\right\vert \rightarrow \infty ,k\in \mathbb{C}%
_{+},  \label{2.81}
\end{equation}
\begin{equation}
u_{s}\left( x,x_{0},k\right) =-\frac{i\exp \left( ik\left\vert
x-x_{0}\right\vert \right) }{16\pi \left\vert x-x_{0}\right\vert k}\left[
\int\limits_{L\left( x,x_{0}\right) }q\left( \xi \right) ds+O\left( \frac{1}{%
k}\right) \right] ,\left\vert k\right\vert \rightarrow \infty ,k\in \mathbb{C%
}_{+},  \label{2.82}
\end{equation}
uniformly for $x\in \overline{G}\cap P_{\omega }\left( x_{0}\right) .$
Therefore, we have proven Lemma 1.

\textbf{Lemma 1}. \emph{The solution }$U\left( x,x_{0},t\right) $\emph{\ of
the problem (\ref{2.1}), (\ref{2.2}) can be represented in the form (\ref%
{2.3}), where the function }$\widetilde{U}\left( x,x_{0},t\right) $\emph{\
satisfies conditions (\ref{2.31})-(\ref{2.6}).\ Furthermore, (\ref{2.7})
holds, where the function }$u\left( x,x_{0},k\right) $ \emph{is the unique
solution of the problem (\ref{1.4}), (\ref{1.50}), (\ref{1.5}) satisfying
conditions (\ref{1.6}), (\ref{1.7}). In addition, for every pair of points }$%
x,x_{0}\in \mathbb{R}^{3}$\emph{\ such that }$x\neq x_{0}$\emph{\ the
function }$u\left( x,x_{0},k\right) $\emph{\ admits the analytic
continuation with respect to }$k$\emph{\ \ from the real line }$\mathbb{R}$ 
\emph{in the half-plane }$\left\{ k\in \mathbb{C}:\func{Im}k>-c_{1}\right\}
, $\emph{\ where }$c_{1}=c_{1}\left( q,\Phi ,x_{0},\omega \right) >0$\emph{\
is the number in (\ref{2.6}). Finally, if }$G\cap P_{\omega }\left(
x_{0}\right) \neq \varnothing ,$ \emph{then asymptotic formulas (\ref{2.81})
and (\ref{2.82}) hold uniformly for }$x\in \overline{G}\cap P_{\omega
}\left( x_{0}\right) $.\emph{\ }

\subsection{Lemmata 2-5}

\label{sec:3.2}

\ \textbf{Lemma 2}. \emph{Let }$x_{0}\in \mathbb{R}^{3}$\emph{\ and }$x\in
G,x\neq x_{0}$\emph{\ be two arbitrary points. Then the function }$u\left(
x,x_{0},k\right) $\emph{\ has at most finite number of zeros in }$\mathbb{C}%
_{+}.$

\textbf{Lemma 3}. \emph{Let }$x_{0}\in \mathbb{R}^{3}$\emph{\ and }$x\in G$%
\emph{\ be two arbitrary points. Assume that} 
\begin{equation*}
\int\limits_{L\left( x,x_{0}\right) }q\left( \xi \right) ds\neq 0.
\end{equation*}
\emph{Then the function }$u_{s}\left( x,x_{0},k\right) $\emph{\ has at most
finite number of zeros in }$\mathbb{C}_{+}.$

Lemmata 2 and 3 follow immediately from (\ref{2.81}) and (\ref{2.82})
respectively.

\textbf{Lemma 4.} \emph{Let }$\gamma >0$\emph{\ be a number.\ Let the
function }$d\left( k\right) $\emph{\ be analytic in }$\left\{ k\in \mathbb{C}%
:\func{Im}k>-\gamma \right\} $\emph{\ and does not have zeros in }$\mathbb{C}%
_{+}.$\emph{\ Assume that } 
\begin{equation*}
d\left( k\right) =\frac{C}{k^{n}}\left[ 1+o\left( 1\right) \right] \exp
\left( ikL\right) ,\left\vert k\right\vert \rightarrow \infty ,k\in \mathbb{C%
}_{+},
\end{equation*}
\emph{where }$C\in \mathbb{C}$\emph{\ and }$n\emph{,}L\in \mathbb{R}$\emph{\
are some numbers and also }$n\geq 0$\emph{. Then the function }$d\left(
k\right) $\emph{\ can be uniquely determined for }$k\in \left\{ k\in \mathbb{%
C}:\func{Im}k>-\gamma \right\} $\emph{\ by the values of }$\left\vert
d\left( k\right) \right\vert $\emph{\ for }$k\in \mathbb{R}$\emph{.
Furthermore, for }$k\in \mathbb{R}$\emph{\ } 
\begin{equation}
\arg d\left( k\right) =\frac{1}{\pi }\lim_{R\rightarrow \infty
}\lim_{\varepsilon \rightarrow 0^{+}}\left[ \int\limits_{-R}^{k-\varepsilon }%
\frac{\ln \left\vert d\left( \xi \right) \right\vert }{k-\xi }d\xi
+\int\limits_{k+\varepsilon }^{R}\frac{\ln \left\vert d\left( \xi \right)
\right\vert }{k-\xi }d\xi \right] +Lk-\arg C+\frac{n\pi }{2}.  \label{4}
\end{equation}

The right hand side of (\ref{4}) can be any of branches of the function $%
\arg $. However, this does not make any difference for us, since we are
interested in the function $d\left( k\right) =\left\vert d\left( k\right)
\right\vert \exp \left[ i\arg d\left( k\right) \right] .$ Since the function 
$d\left( k\right) $ is uniquely determined on the real line, then the
analyticity of this function implies that it is uniquely determined in $%
\left\{ k\in \mathbb{C}:\func{Im}k>-\gamma \right\} .$ Lemma 4 follows
immediately from Proposition 4.2 of \cite{Kl11}. Hence, we omit the proof.

\textbf{Lemma 5}. \emph{Let the function }$d\left( k\right) $\emph{\ be
analytic for all }$k\in \mathbb{R}.$ \emph{Then the function }$\left\vert
d\left( k\right) \right\vert $\emph{\ can be uniquely for all }$k\in \mathbb{%
R}$ \emph{by the values of }$\left\vert d\left( k\right) \right\vert $ \emph{%
\ for }$k\in \left( a,b\right) $\emph{.}

This lemma was actually proven in subsection 1.1 for the function $r\left(
k\right) $.

\subsection{Proof of Theorem 1}

\label{sec:3.3}

Choose an arbitrary point $x_{0}\in S$ and an arbitrary point $x\in
B_{\varepsilon }\left( x_{0}\right) $ such that $x\neq x_{0}.$ Denote 
\begin{equation}
h_{1}\left( k\right) =u_{1}\left( x,x_{0},k\right) ,h_{2}\left( k\right)
=u_{2}\left( x,x_{0},k\right) .  \label{3}
\end{equation}
It follows from Lemma 1 that we can regard $h_{1}\left( k\right) $ and $%
h_{2}\left( k\right) $ as analytic functions in the half-plane $\left\{ k\in 
\mathbb{C}:\func{Im}k>-\theta _{1}\right\} ,$ where $\theta _{1}>0$ is a
certain number. Next, since $\left\vert h_{1}\left( k\right) \right\vert
=\left\vert h_{2}\left( k\right) \right\vert $ for $k\in \left( a,b\right) ,$
then Lemma 5 implies that 
\begin{equation}
\left\vert h_{1}\left( k\right) \right\vert =\left\vert h_{2}\left( k\right)
\right\vert ,\forall k\in \mathbb{R}.  \label{5}
\end{equation}
By Lemma 2 each of functions $h_{1}\left( k\right) ,h_{2}\left( k\right) $
has at most finite number of zeros in $\mathbb{C}_{+}.$ Let $\left\{
a_{1},...,a_{n}\right\} \subset \left( \mathbb{C}_{+}\diagdown \mathbb{R}%
\right) $ and $\left\{ b_{1},...,b_{m}\right\} \subset \left( \mathbb{C}%
_{+}\diagdown \mathbb{R}\right) $ be sets of all zeros of functions $%
h_{1}\left( k\right) $ and $h_{2}\left( k\right) $ respectively in the upper
half-plane. Here and below each zero is counted as many times as its order
is. Let $\left\{ a_{1}^{\prime },...,a_{r_{1}}^{\prime }\right\} \subset 
\mathbb{R}$ and $\left\{ b_{1}^{\prime },...,b_{r_{2}}^{\prime }\right\}
\subset \mathbb{R}$ be sets of all real zeros of functions $h_{1}\left(
k\right) $ and $h_{2}\left( k\right) $ respectively.

First, we prove that 
\begin{equation}
\left\{ a_{1}^{\prime },...,a_{r_{1}}^{\prime }\right\} =\left\{
b_{1}^{\prime },...,b_{r_{2}}^{\prime }\right\} .  \label{4.0}
\end{equation}
Let, for example $a_{1}^{\prime }$ be the zero of the order $n_{1}\geq 1$ of
the function $h_{1}\left( k\right) $ as well as the zero of the order $%
m_{1}\geq 0$ of the function $h_{2}\left( k\right) .$ Then $h_{1}\left(
k\right) =\left( k-a_{1}^{\prime }\right) ^{n_{1}}\widehat{h}_{1}\left(
k\right) $ and $h_{2}\left( k\right) =\left( k-a_{1}^{\prime }\right)
^{m_{1}}\widehat{h}_{1}\left( k\right) ,$ where 
\begin{equation}
\widehat{h}_{1}\left( a_{1}^{\prime }\right) \widehat{h}_{2}\left(
a_{1}^{\prime }\right) \neq 0.  \label{4.01}
\end{equation}
Using (\ref{5}), we obtain 
\begin{equation}
\left\vert \left( k-a_{1}^{\prime }\right) ^{n_{1}}\right\vert \cdot
\left\vert \widehat{h}_{1}\left( k\right) \right\vert =\left\vert \left(
k-a_{1}^{\prime }\right) ^{m_{1}}\right\vert \cdot \left\vert \widehat{h}%
_{2}\left( k\right) \right\vert ,\text{ }\forall k\in \mathbb{R}.
\label{4.02}
\end{equation}
Assume, for example that $m_{1}<n_{1}.$ Dividing both sides of (\ref{4.02})
by $\left\vert \left( k-a_{1}^{\prime }\right) ^{m_{1}}\right\vert $ and
setting the limit at $k\rightarrow a_{1}^{\prime },$ we obtain $\left\vert 
\widehat{h}_{2}\left( a_{1}^{\prime }\right) \right\vert =0.$ This
contradicts to (\ref{4.01}). Hence, $m_{1}=n_{1}.$ Since $a_{1}^{\prime }$
is an arbitrary element of the set $\left\{ a_{1}^{\prime
},...,a_{r_{1}}^{\prime }\right\} ,$ then (\ref{4.0}) follows.

Let $\left\{ c_{1},...,c_{r}\right\} \subset \mathbb{R}$ be the set of all
real zeros of both functions $h_{1}\left( k\right) $ and $h_{2}\left(
k\right) .$ Define functions $\widetilde{h}_{1}\left( k\right) ,\widetilde{h}%
_{2}\left( k\right) $ as 
\begin{equation}
\widetilde{h}_{1}\left( k\right) =h_{1}\left( k\right) \left(
\prod\limits_{j=1}^{n}\frac{k-\overline{a}_{j}}{k-a_{j}}\right) \left(
\prod\limits_{s=1}^{r}\frac{1}{k-c_{s}}\right) ,  \label{4.03}
\end{equation}
\begin{equation}
\widetilde{h}_{2}\left( k\right) =h_{2}\left( k\right) \left(
\prod\limits_{j=1}^{m}\frac{k-\overline{b}_{j}}{k-b_{j}}\right) \left(
\prod\limits_{s=1}^{r}\frac{1}{k-c_{s}}\right) .  \label{4.04}
\end{equation}
Hence, $\widetilde{h}_{1}\left( k\right) $ and $\widetilde{h}_{2}\left(
k\right) $ are analytic functions in $\left\{ k\in \mathbb{C}:\func{Im}%
k>-\theta _{1}\right\} .$ Also, these functions do not have zeros in $%
\mathbb{C}_{+}.$ Furthermore, it follows from (\ref{5}), (\ref{4.03}) and (%
\ref{4.04}) and $\left\vert \widetilde{h}_{1}\left( k\right) \right\vert
=\left\vert \widetilde{h}_{2}\left( k\right) \right\vert ,\forall k\in 
\mathbb{R}.$ Also, using (\ref{2.81}), we obtain the following asymptotic
behavior of both functions $\widetilde{h}_{1}\left( k\right) $ and $%
\widetilde{h}_{2}\left( k\right) $ 
\begin{equation*}
\widetilde{h}_{j}\left( k\right) =\frac{\exp \left( ik\left\vert
x-x_{0}\right\vert \right) }{4\pi \left\vert x-x_{0}\right\vert k^{r}}\left[
1+O\left( \frac{1}{k}\right) \right] ,\left\vert k\right\vert \rightarrow
\infty ,k\in \mathbb{C}_{+},j=1,2.
\end{equation*}
Hence, Lemma 4 implies that $\widetilde{h}_{1}\left( k\right) =\widetilde{h}%
_{2}\left( k\right) ,\forall k\in \mathbb{R}.$ Hence, (\ref{4.03}) and (\ref%
{4.04}) lead to 
\begin{equation*}
h_{1}\left( k\right) \prod\limits_{j=1}^{n}\frac{k-\overline{a}_{j}}{k-a_{j}}%
=h_{2}\left( k\right) \prod\limits_{j=1}^{m}\frac{k-\overline{b}_{j}}{k-b_{j}%
},\text{ }\forall k\in \mathbb{R}.
\end{equation*}
Hence, 
\begin{equation}
h_{1}\left( k\right) \prod\limits_{j=1}^{m}\frac{k-b_{j}}{k-\overline{b}_{j}}%
=h_{2}\left( k\right) \prod\limits_{j=1}^{n}\frac{k-a_{j}}{k-\overline{a}_{j}%
},\text{ }\forall k\in \mathbb{R}.  \label{4.1}
\end{equation}
Rewrite (\ref{4.1}) as 
\begin{equation}
h_{1}\left( k\right) +h_{1}\left( k\right) \left( \prod\limits_{j=1}^{m}%
\frac{k-b_{j}}{k-\overline{b}_{j}}-1\right) =h_{2}\left( k\right)
+h_{2}\left( k\right) \left( \prod\limits_{j=1}^{n}\frac{k-a_{j}}{k-%
\overline{a}_{j}}-1\right) .  \label{4.2}
\end{equation}
Consider the function $w_{1}\left( k\right) ,$%
\begin{equation*}
w_{1}\left( k\right) =\prod\limits_{j=1}^{n}\frac{k-a_{j}}{k-\overline{a}_{j}%
}-1.
\end{equation*}
This function can be rewritten in the form 
\begin{equation*}
w_{1}\left( k\right) =Q\left( k\right) \prod\limits_{j=1}^{n}\frac{1}{k-%
\overline{a}_{j}},
\end{equation*}
where $Q\left( k\right) $ is a polynomial of the degree less than $n$. By a
partial fraction expansion, $A\left( k\right) $ can be written in the form 
\begin{equation*}
w_{1}\left( k\right) =\sum\limits_{j=1}^{n^{\prime }}\frac{C_{j}}{\left( k-%
\overline{a}_{j}\right) ^{s_{j}}},
\end{equation*}
where $C_{j}\in \mathbb{C}$ are some numbers and $s_{j},n^{\prime }\geq 1$
are integers. Direct calculations verify that the inverse Fourier transform $%
\mathcal{F}^{-1}$ of the function $\left( k-\overline{a}_{j}\right)
^{-s_{j}} $ is 
\begin{equation*}
\mathcal{F}^{-1}\left( \frac{1}{\left( k-\overline{a}_{j}\right) ^{s_{j}}}%
\right) =H\left( t\right) \widehat{C}_{j}t^{s_{j}-1}\exp \left( -i\overline{a%
}_{j}t\right)
\end{equation*}
with a certain constant $\widehat{C}_{j}\in \mathbb{C}.$ Hence, 
\begin{equation}
\mathcal{F}^{-1}\left( w_{1}\left( k\right) \right) :=\lambda _{1}\left(
t\right) =H\left( t\right) \sum\limits_{j=1}^{n^{\prime }}\widetilde{C}%
_{j}t^{s_{j}-1}\exp \left( -i\overline{a}_{j}t\right) ,  \label{4.3}
\end{equation}
where constants $\widetilde{C}_{j}=C_{j}\widehat{C}_{j}\in \mathbb{C}.$
Similarly, denoting 
\begin{equation*}
w_{2}\left( k\right) =\prod\limits_{j=1}^{m}\frac{k-b_{j}}{k-\overline{b}_{j}%
}-1,
\end{equation*}
we obtain 
\begin{equation}
\mathcal{F}^{-1}\left( w_{2}\left( k\right) \right) :=\lambda _{2}\left(
t\right) =H\left( t\right) \sum\limits_{j=1}^{m^{\prime
}}B_{j}t^{s_{j}-1}\exp \left( -i\overline{b}_{j}t\right) ,  \label{4.4}
\end{equation}
where constants $B_{j}\in \mathbb{C}.$ For $j=1,2$ let $\mathcal{F}%
^{-1}\left( h_{j}\right) =\widehat{h}_{j}\left( t\right) .$ By (\ref{2.7})
and (\ref{3}) 
\begin{equation}
\widehat{h}_{j}\left( t\right) =U_{j}\left( x,x_{0},t\right) ,  \label{4.41}
\end{equation}
where $U_{j}\left( x,x_{0},t\right) $ is the solution of the problem (\ref%
{2.1}), (\ref{2.2}) with $q\left( x\right) =q_{j}\left( x\right) .$ We now
apply the operator $\mathcal{F}^{-1}$ to both sides of (\ref{4.2}).\ Using (%
\ref{4.3}), (\ref{4.4}) and the convolution theorem, we obtain 
\begin{equation}
\widehat{h}_{1}\left( t\right) +\int\limits_{0}^{t}\widehat{h}_{1}\left(
t-s\right) \lambda _{2}\left( s\right) ds=\widehat{h}_{2}\left( t\right)
+\int\limits_{0}^{t}\widehat{h}_{2}\left( t-s\right) \lambda _{1}\left(
s\right) ds.  \label{4.5}
\end{equation}

Since $x_{0}\in S,x\in B_{\varepsilon }\left( x_{0}\right) ,$ then (\ref{1}%
), (\ref{2}) and (\ref{2.32}) imply that 
\begin{equation}
D\left( x,x_{0},t\right) \subset \left( G\diagdown \overline{\Omega }\right)
,\forall t\in \left( \left\vert x-x_{0}\right\vert ,\varepsilon \right) .
\label{4.7}
\end{equation}
Since $q_{1}\left( x\right) =q_{2}\left( x\right) $ for $x\in \mathbb{R}%
^{3}\diagdown \Omega ,$ then (\ref{2.401}), (\ref{2.402}), (\ref{2.403}) and
(\ref{4.7}) imply that 
\begin{equation}
U_{1}\left( x,x_{0},t\right) =U_{2}\left( x,x_{0},t\right) =U\left(
x,x_{0},t\right) ,\forall t\in \left( \left\vert x-x_{0}\right\vert
,\varepsilon \right) .  \label{4.8}
\end{equation}
By (\ref{4.41}) and (\ref{4.8}) 
\begin{equation}
\widehat{h}_{1}\left( t\right) =\widehat{h}_{2}\left( t\right) =\widehat{h}%
\left( t\right) =U\left( x,x_{0},t\right) \text{, }\forall t\in \left(
\left\vert x-x_{0}\right\vert ,\varepsilon \right) .  \label{4.81}
\end{equation}
Hence, (\ref{4.5}) implies that 
\begin{equation}
\int\limits_{0}^{t}\widehat{h}\left( t-s\right) \lambda \left( s\right) ds=0,%
\text{ }\forall t\in \left( \left\vert x-x_{0}\right\vert ,\varepsilon
\right) ,  \label{4.9}
\end{equation}
where 
\begin{equation}
\lambda \left( t\right) =\lambda _{1}\left( t\right) -\lambda _{2}\left(
t\right) .  \label{4.10}
\end{equation}

Using (\ref{2.3}), (\ref{2.33}), (\ref{2.4}), (\ref{2.41}) and (\ref{4.81}),
we obtain 
\begin{equation}
\widehat{h}\left( t\right) =\frac{\delta \left( t-\left\vert
x-x_{0}\right\vert \right) }{4\pi \left\vert x-x_{0}\right\vert }+p\left(
t\right) ,  \label{4.11}
\end{equation}
\begin{equation}
p\left( t\right) =0,t\in \left( 0,\left\vert x-x_{0}\right\vert \right) ,
\label{4.12}
\end{equation}
\begin{equation}
\lim_{t\rightarrow \left\vert x-x_{0}\right\vert ^{+}}p\left( t\right) =-%
\frac{1}{16\pi \left\vert x-x_{0}\right\vert }\int\limits_{L\left(
x,x_{0}\right) }q\left( \xi \right) ds,  \label{4.121}
\end{equation}
\begin{equation}
p\in C^{2}\left( t\geq \left\vert x-x_{0}\right\vert \right) .  \label{4.14}
\end{equation}
Introduce a new variable $\tau \Leftrightarrow t,$ where $\tau =t-\left\vert
x-x_{0}\right\vert .$ Then (\ref{4.9}) and (\ref{4.11})-(\ref{4.14}) lead to
the following integral equation of the Volterra type with the continuous
kernel $p\left( \tau +\left\vert y-y_{0}\right\vert -s\right) $ and with
respect to the function $\lambda \left( \tau \right) $ 
\begin{equation}
\lambda \left( \tau \right) +4\pi \left\vert x-x_{0}\right\vert
\int\limits_{0}^{\tau }p\left( \tau +\left\vert x-x_{0}\right\vert -s\right)
\lambda \left( s\right) ds=0,\text{ }\forall \tau \in \left( 0,\varepsilon
-\left\vert x-x_{0}\right\vert \right) .  \label{4.15}
\end{equation}
Hence, $\lambda \left( \tau \right) =0$ for all $\tau \in \left(
0,\varepsilon -\left\vert x-x_{0}\right\vert \right) .$ On the other hand, (%
\ref{4.3}), (\ref{4.4}) and (\ref{4.10}) imply that $\lambda \left( t\right) 
$ is analytic function of the real variable $t>0.$\ Hence, $\lambda \left(
t\right) =0,$ $\forall t\geq 0.$ This implies that $\left\{
a_{1},...,a_{n}\right\} =\left\{ b_{1},...,b_{m}\right\} .$ Thus, (\ref{3})
and (\ref{4.1}) lead to $u_{1}\left( x,x_{0},k\right) =u_{2}\left(
x,x_{0},k\right) ,\forall k\in \mathbb{R}.$

Since $x_{0}\in S$ and $x\in B_{\varepsilon }\left( x_{0}\right) $ are two
arbitrary points such that $x\neq x_{0}$, then we have established that 
\begin{equation}
u_{1}\left( x,x_{0},k\right) =u_{2}\left( x,x_{0},k\right) ,\forall x_{0}\in
S,\forall x\in B_{\varepsilon }\left( x_{0}\right) ,x\neq x_{0},\forall k\in 
\mathbb{R}.  \label{6.1}
\end{equation}
Consider an arbitrary point $y_{0}\in S.$ Since $q_{1}\left( x\right)
=q_{2}\left( x\right) =q\left( x\right) $ in $\mathbb{R}^{3}\diagdown \Omega
,$ then, using (\ref{1.4}), we obtain 
\begin{equation*}
\Delta _{x}u_{j}+k^{2}u_{j}-q\left( x\right) u_{j}=-\delta \left(
x-y_{0}\right) ,x\in \mathbb{R}^{3}\diagdown \Omega ,j=1,2.
\end{equation*}
Hence, (\ref{6.1}) and the well known result about uniqueness of the
continuation problem for elliptic equations (see, e.g. \S 1 of chapter 4 of 
\cite{LRS}) imply that $u_{1}\left( x,y_{0},k\right) =u_{2}\left(
x,y_{0},k\right) ,\forall x\in \mathbb{R}^{3}\diagdown \Omega ,\forall k\in 
\mathbb{R}.$ Hence, 
\begin{equation}
u_{1}\left( x,x_{0},k\right) =u_{2}\left( x,x_{0},k\right) ,\forall
x,x_{0}\in S,x\neq x_{0},\forall k\in \mathbb{R}.  \label{6.2}
\end{equation}
Using (\ref{1.6}), (\ref{1.7}), (\ref{2.81}) and (\ref{2.82}), we obtain for 
$j=1,2$ 
\begin{equation}
\lim_{k\rightarrow \infty }\left[ 4ik\left( \frac{u_{j}}{u_{0}}-1\right)
\left( x,x_{0},k\right) \right] =\int\limits_{L\left( x,x_{0}\right)
}q_{j}\left( \xi \right) ds,\text{ }\forall x,x_{0}\in \mathbb{R}^{3},x\neq
x_{0}.  \label{6.3}
\end{equation}
Hence, (\ref{6.2}) and (\ref{6.3}) lead to 
\begin{equation}
\int\limits_{L\left( x,x_{0}\right) }\left( q_{1}-q_{2}\right) \left( \xi
\right) ds=0,\text{ }\forall x,x_{0}\in S.  \label{6.4}
\end{equation}
Finally, (\ref{6.4}) and the classical uniqueness theorem for the Radon
transform implies that $q_{1}\left( x\right) \equiv q_{2}\left( x\right) .$ $%
\square $

The idea of using (\ref{6.4}) for the proof of the uniqueness of an inverse
scattering problem in the time domain can be found in \S 1 of chapter 7 of 
\cite{LRS}

\section{Proof of Theorem 2}

\label{sec.4}

Consider an arbitrary point $x_{0}\in S.$ Since $q_{1}\left( x\right)
=q_{2}\left( x\right) =q\left( x\right) $\emph{\ }for\emph{\ }$x\in \mathbb{R%
}^{3}\diagdown \Omega $ and $q\left( x\right) \neq 0,\forall x\in S$, then (%
\ref{1}) and (\ref{2}) imply that one can choose $\varepsilon >0$ so small
that 
\begin{equation}
\int\limits_{L\left( y,x_{0}\right) }q_{1}\left( \xi \right)
ds=\int\limits_{L\left( y,x_{0}\right) }q_{2}\left( \xi \right)
ds=\int\limits_{L\left( y,x_{0}\right) }q\left( \xi \right) ds\neq 0,\forall
y\in B_{\varepsilon }\left( x_{0}\right) .  \label{4.16}
\end{equation}
Choose an arbitrary point $x\in B_{\varepsilon }\left( x_{0}\right) $ such
that $x\neq x_{0}.$ Denote $h_{s,j}\left( k\right) =u_{s,j}\left(
x,x_{0},k\right) .$ It follows from (\ref{4.16}) and Lemmata 1, 3, 4 and 5
that we can apply to functions $h_{s,j}\left( k\right) $ the same procedure
as the one described in section 3.3. For brevity we keep notations (\ref{4.3}%
), (\ref{4.4}) and (\ref{4.10}). For $j=1,2$ let $\widehat{h}_{s,j}\left(
t\right) =\mathcal{F}^{-1}\left( h_{s,j}\right) .$ Let $\widetilde{U}\left(
x,x_{0},t\right) $ be the function defined in (\ref{2.3}), (\ref{2.31}).
Then, using (\ref{4.11}), (\ref{4.12}) and (\ref{4.14}), we obtain
analogously with (\ref{4.81}) 
\begin{equation}
\widehat{h}_{s,j}\left( t\right) =\widetilde{U}\left( x,x_{0},t\right)
=p\left( t\right) ,\forall t\in \left( \left\vert x-x_{0}\right\vert
,\varepsilon \right) ,j=1,2.  \label{4.17}
\end{equation}
Hence, using (\ref{4.17}), we obtain similarly with (\ref{4.15}) 
\begin{equation}
\int\limits_{0}^{\tau }p\left( \tau -s+\left\vert x-x_{0}\right\vert \right)
\lambda \left( s\right) ds=0,\forall \tau \in \left( 0,\varepsilon
-\left\vert x-x_{0}\right\vert \right) ,  \label{4.171}
\end{equation}
where $\tau =t-\left\vert x-x_{0}\right\vert .$ Differentiating both sides
of (\ref{4.171}) with respect to $\tau $ and using (\ref{4.121}), (\ref{4.14}%
) and (\ref{4.16}), we obtain 
\begin{equation*}
\lambda \left( \tau \right) -m\left( x,x_{0}\right) \int\limits_{0}^{\tau
}p^{\prime }\left( \tau +\left\vert x-x_{0}\right\vert -s\right) \lambda
\left( s\right) ds=0,\text{ }\forall \tau \in \left( 0,\varepsilon
-\left\vert x-x_{0}\right\vert \right) ,
\end{equation*}
\begin{equation*}
m\left( x,x_{0}\right) =\frac{16\pi }{\left\vert x-x_{0}\right\vert }\left(
\int\limits_{L\left( x,x_{0}\right) }q\left( \xi \right) ds\right) ^{-1}.
\end{equation*}
Hence, similarly with the proof of Theorem 1, we conclude that $\lambda
\left( t\right) =0,\forall t\geq 0.$ This leads to 
\begin{equation}
u_{s,1}\left( x,x_{0},k\right) =u_{s,2}\left( x,x_{0},k\right) ,\forall k\in 
\mathbb{R}.  \label{4.50}
\end{equation}
Next, since $u_{j}\left( x,x_{0},k\right) =u_{0}\left( x,x_{0},k\right)
+u_{s,j}\left( x,x_{0},k\right) ,j=1,2$ and since again $x_{0}\in S$ and $%
x\in B_{\varepsilon }\left( x_{0}\right) $ are two arbitrary points such
that $x\neq x_{0},$ then (\ref{4.50}) implies (\ref{6.1}). The rest of the
proof is the same as the proof of Theorem 1 after (\ref{6.1}). $\square $

\section{Proofs of Theorems 3 and 4}

\label{sec.5}

\subsection{Functions $V$ and $v$}

\label{sec.5.1}

Consider the following Cauchy problem 
\begin{equation}
V_{tt}=\Delta V-q\left( x\right) V,\text{ }\left( x,t\right) \in \mathbb{R}%
^{3}\times \left( 0,\infty \right) ,  \label{2.9}
\end{equation}
\begin{equation}
V\left( x,0\right) =0,V_{t}\left( x,0\right) =g\left( x\right) .
\label{2.10}
\end{equation}
Corollary 4.2 of Chapter 4 of the book \cite{Lad} implies that there exists
unique solution $V\in H^{2}\left( \mathbb{R}^{3}\times \left( 0,T\right)
\right) ,\forall T>0$ of the problem (\ref{2.9}), (\ref{2.10}). By the
Kirchhoff formula this function $V\left( x,t\right) $ is the solution of the
following integral equation 
\begin{equation}
V\left( x,t\right) =\frac{1}{4\pi t}\int\limits_{\left\vert x-\xi
\right\vert =t}g\left( \xi \right) dS_{\xi }-\int\limits_{\left\vert x-\xi
\right\vert <t}\frac{q\left( \xi \right) V\left( \xi ,t-\left\vert x-\xi
\right\vert \right) }{4\pi \left\vert x-\xi \right\vert }d\xi .
\label{2.101}
\end{equation}
Construct functions $V_{n}\left( x,t\right) $ as 
\begin{equation}
V_{0}\left( x,t\right) =\frac{1}{4\pi }\int\limits_{\left\vert x-\xi
\right\vert =t}g\left( \xi \right) dS_{\xi },  \label{2.100}
\end{equation}
\begin{equation}
V_{n}\left( x,t\right) =-\int\limits_{\left\vert x-\xi \right\vert <t}\frac{%
q\left( \xi \right) V_{n-1}\left( \xi ,t-\left\vert x-\xi \right\vert
\right) }{4\pi \left\vert x-\xi \right\vert }d\xi ,\text{ }n=1,2,...,
\label{2.102}
\end{equation}
The above mentioned technique of \S 3 of chapter 2 of the book \cite{Rom1}
implies that the function $V\left( x,t\right) $ can be represented as 
\begin{equation}
V\left( x,t\right) =\sum\limits_{n=0}^{\infty }V_{n}\left( x,t\right) ,
\label{2.11}
\end{equation}
and this series converges in the norm of the space $C^{4}\left( \overline{%
\Phi }\times \left[ 0,T\right] \right) $ for any bounded domain $\Phi
\subset \mathbb{R}^{3}$ and for any number $T>0.$ Hence, 
\begin{equation}
V\in C^{4}\left( \overline{\Phi }\times \left[ 0,T\right] \right) .
\label{2.12}
\end{equation}

Using again Lemma 6 in Chapter 10 of the book \cite{V} as well as Remark 3
after that lemma, we obtain that for any bounded domain $\Phi \subset 
\mathbb{R}^{3}$ there exist constants $C_{2}=C_{2}\left( q,g,\Phi \right)
>0,c_{2}=c_{2}\left( q,g,\Phi \right) >0$ depending only on functions $q,g$
and the domain $\Phi $ such that 
\begin{equation}
\left\vert D_{x,t}^{\beta }V\left( x,t\right) \right\vert \leq
C_{2}e^{-c_{2}t},\text{ }\forall x\in \Phi ,\forall t\geq 0;\left\vert \beta
\right\vert =0,1,...,4.  \label{2.13}
\end{equation}
Hence, using again Theorem 6 of Chapter 9 of \cite{V}, Theorem 3.3 of \cite%
{V1} and Theorem 6.17 of \cite{GT}, we obtain that the Fourier transform of
the function $V\left( x,t\right) $ is the unique solution $v\left(
x,k\right) \in C^{5+\alpha }\left( \mathbb{R}^{3}\right) ,\forall \alpha \in
\left( 0,1\right) $ of the problem (\ref{1.12})-(\ref{1.13}), i.e. 
\begin{equation}
v\left( x,k\right) =\mathcal{F}\left( V\right) ,\forall x\in \mathbb{R}%
^{3},\forall k\in \mathbb{R}.  \label{2.14}
\end{equation}
Furthermore, it follows from (\ref{2.13}) and (\ref{2.14}) that for every
point $x\in G$ the function $v\left( x,k\right) $ admits the analytic
continuation with respect to $k$ from the real line $\mathbb{R}$ in the
half-plane $\left\{ k\in \mathbb{C}:\func{Im}k>-c_{2}\left( q,g,G\right)
\right\} .$

Hence, using the integration by parts in integral (\ref{2.14}) of the
Fourier transform as well as (\ref{2.10}), (\ref{2.12}) and (\ref{2.13}), we
obtain the following asymptotic behavior of the function $v\left( x,k\right)
,$ uniformly for $x\in \overline{G}$ 
\begin{equation}
v\left( x,k\right) =\frac{1}{k^{2}}\left[ -g\left( x\right) +O\left( \frac{1%
}{k}\right) \right] ,\left\vert k\right\vert \rightarrow \infty ,k\in 
\mathbb{C}_{+}.  \label{2.15}
\end{equation}
Let 
\begin{equation}
V_{s}\left( x,t\right) =V\left( x,t\right) -V_{0}\left( x,t\right) .
\label{2.150}
\end{equation}
Then 
\begin{equation}
v_{s}\left( x,k\right) =\mathcal{F}\left( V_{s}\right) .  \label{2.151}
\end{equation}
Next, using (\ref{2.9}) and (\ref{2.10}), we obtain 
\begin{equation}
\partial _{t}^{2}V_{s}=\Delta V_{s}-q\left( x\right) \left(
V_{s}+V_{0}\right) ,\left( x,t\right) \in \mathbb{R}^{3}\times \left(
0,\infty \right) ,  \label{2.152}
\end{equation}
\begin{equation}
V_{s}\left( x,0\right) =\partial _{t}V_{s}\left( x,0\right) =0.
\label{2.153}
\end{equation}
Hence, (\ref{2.10}), (\ref{2.152}) and (\ref{2.153}) imply that 
\begin{equation}
\partial _{t}^{r}V_{s}\left( x,0\right) =0\text{ for }r=0,1,2\text{ and }%
\partial _{t}^{3}V_{s}\left( x,0\right) =-\left( qg\right) \left( x\right) .
\label{2.154}
\end{equation}
Hence, using (\ref{2.154}) and the integration by parts in the right hand
side of (\ref{2.151}), we obtain the following asymptotic behavior of the
function $v_{s}\left( x,k\right) ,$ uniformly for $x\in \overline{G}$ 
\begin{equation}
v_{s}\left( x,k\right) =\frac{1}{k^{4}}\left[ -\left( qg\right) \left(
x\right) +o\left( 1\right) \right] ,\left\vert k\right\vert \rightarrow
\infty ,k\in \mathbb{C}_{+}.  \label{2.16}
\end{equation}

Thus, we have proven Lemma 6.

\textbf{Lemma 6}. \emph{There exists unique solution }$V\left( x,t\right) $%
\emph{\ of the problem (\ref{2.9}), (\ref{2.10}) such that (\ref{2.12}) is
valid for every bounded domain }$\Phi \subset \mathbb{R}^{3}$\emph{\ and for
every }$T>0.$\emph{\ Estimate (\ref{2.13}) is valid for this function }$%
V\left( x,t\right) .$\emph{\ In addition, the Fourier transform (\ref{2.14})
of the function }$V\left( x,t\right) $\emph{\ is the unique solution }$%
v\left( x,k\right) \in C^{5+\alpha }\left( \mathbb{R}^{3}\right) ,\forall
\alpha \in \left( 0,1\right) $\emph{\ of the problem (\ref{1.12})-(\ref{1.13}%
). Also, for every }$x\in \Phi $ \emph{the function }$v\left( x,k\right) $ 
\emph{admits the analytic continuation with respect to }$k$\emph{\ from the
real line }$\mathbb{R}$ \emph{in the half-plane }$\left\{ k\in C:\func{Im}%
k>-c_{2}\right\} .$ \emph{Finally, asymptotic formulas (\ref{2.15}) and (\ref%
{2.16}) hold uniformly for }$x\in \overline{G}$\emph{.}

Lemma 7 follows immediately from Lemma 6, (\ref{1.11}), (\ref{2.15}) and (%
\ref{2.16}).

\textbf{Lemma 7}. \emph{For every point }$x\in \overline{G}_{1}$\emph{\
there exists at most finite number of zeros of the function }$v\left(
x,k\right) $\emph{\ in }$\mathbb{C}_{+}.$ \emph{Next, assume that there
exists a point }$x^{\prime }\in \overline{G}_{1}$\emph{\ such that }$q\left(
x^{\prime }\right) \neq 0.$\emph{\ Then the function }$v_{s}\left( x^{\prime
},k\right) $\emph{\ has at most finite number of zeros in }$\mathbb{C}_{+}.$

\subsection{Proof of Theorem 3}

\label{sec.5.2}

Consider an arbitrary point $\widetilde{x}\in S$ and denote similarly with (%
\ref{3}) $h_{j}\left( k\right) =v_{j}\left( \widetilde{x},k\right) ,j=1,2.$
By Lemma 6 there exists a number $\theta _{2}>0$ such that each of functions 
$h_{1}\left( k\right) $, $h_{2}\left( k\right) $ admits the analytic
continuation from the real line $\mathbb{R}$ in the half-plane $\left\{ k\in 
\mathbb{C}:\func{Im}k>-\theta _{2}\right\} .$ By Lemma 7 each function $%
h_{1}\left( k\right) ,h_{2}\left( k\right) $ has at most finite number of
zeros in $\mathbb{C}_{+}.$ Hence, the asymptotic behavior (\ref{2.15})
enables us to apply the technique of section 3.3. For $j=1,2$ let $\widehat{h%
}_{j}\left( t\right) =\mathcal{F}^{-1}\left( h_{j}\right) .$ Let the
function $V_{j}\left( x,t\right) $ satisfying (\ref{2.12}) be the solution
of the problem (\ref{2.9}), (\ref{2.10}) with $q\left( x\right)
:=q_{j}\left( x\right) $. Then (\ref{2.14}) implies that 
\begin{equation}
\widehat{h}_{j}\left( t\right) =V_{j}\left( \widetilde{x},t\right) ,j=1,2.
\label{4.18}
\end{equation}
Since $\widetilde{x}\in S,$ then it follows from (\ref{1}) and (\ref{2.101}%
)-(\ref{2.11}) that 
\begin{equation}
V_{1}\left( \widetilde{x},t\right) =V_{2}\left( \widetilde{x},t\right)
=V\left( \widetilde{x},t\right) ,\text{ }\forall t\in \left( 0,\varepsilon
\right) .  \label{4.181}
\end{equation}
We briefly note that another way of establishing (\ref{4.181}) is via the
energy estimate. Using (\ref{4.18}) and (\ref{4.181}), we obtain 
\begin{equation}
\widehat{h}_{1}\left( t\right) =\widehat{h}_{2}\left( t\right) =V\left( 
\widetilde{x},t\right) ,\text{ }\forall t\in \left( 0,\varepsilon \right) .
\label{4.19}
\end{equation}
Using arguments, which are completely analogous with those of section 3.3,
keeping the same notations (\ref{4.3}), (\ref{4.4}), (\ref{4.10}) and using (%
\ref{4.19}), we obtain the following analog of (\ref{4.171}) 
\begin{equation}
\int\limits_{0}^{t}V\left( \widetilde{x},t-\tau \right) \lambda \left( \tau
\right) d\tau =0,\text{ }\forall t\in \left( 0,\varepsilon \right) .
\label{4.190}
\end{equation}
Differentiating both sides of (\ref{4.190}) twice, using initial conditions (%
\ref{2.10}) as well as (\ref{2.12}), we obtain the following Volterra
integral equation of the second kind with respect to the function $c\left(
t\right) $%
\begin{equation}
\lambda \left( t\right) +\frac{1}{g\left( \widetilde{x}\right) }%
\int\limits_{0}^{t}V_{tt}\left( \widetilde{x},t-\tau \right) \lambda \left(
\tau \right) d\tau =0,\text{ }\forall t\in \left( 0,\varepsilon \right) .
\label{4.191}
\end{equation}
Hence, analogously with subsection 3.3, we conclude that $\lambda \left(
t\right) =0,$ $\forall t\geq 0$ and 
\begin{equation}
v_{1}\left( \widetilde{x},k\right) =v_{2}\left( \widetilde{x},k\right)
,\forall k\in \mathbb{R}.  \label{4.20}
\end{equation}

Since $\widetilde{x}\in S$ is an arbitrary point, then (\ref{4.20}) implies
that 
\begin{equation}
v_{1}\left( x,k\right) =v_{2}\left( x,k\right) ,\forall x\in S,\forall k\in 
\mathbb{R}.  \label{4.21}
\end{equation}
Since $V_{j}\left( x,t\right) =\mathcal{F}^{-1}\left( v_{j}\right) ,j=1,2,$
then (\ref{4.21}) leads to 
\begin{equation}
V_{1}\left( x,t\right) \mid _{S_{\infty }}=V_{2}\left( x,t\right) \mid
_{S_{\infty }}:=\eta \left( x,t\right) ,  \label{4.30}
\end{equation}
where $S_{\infty }=S\times \left( 0,\infty \right) .$ Let $\widehat{V}\left(
x,t\right) $ be any of two functions $V_{1}\left( x,t\right) ,V_{2}\left(
x,t\right) .$ Since 
\begin{equation}
q_{1}\left( x\right) =q_{2}\left( x\right) =q\left( x\right) \text{ for }%
x\in \mathbb{R}^{3}\diagdown \Omega ,  \label{4.211}
\end{equation}
then (\ref{2.9}), (\ref{2.10}) and (\ref{4.30}) imply that the function $%
\widehat{V}\left( x,t\right) $ satisfies the following conditions 
\begin{equation}
\widehat{V}_{tt}=\Delta \widehat{V}-q\left( x\right) \widehat{V},\text{ }%
\left( x,t\right) \in \left( \mathbb{R}^{3}\diagdown G_{1}\right) \times
\left( 0,\infty \right) ,  \label{4.31}
\end{equation}
\begin{equation}
\widehat{V}\left( x,0\right) =0,\partial _{t}\widehat{V}\left( x,0\right)
=g\left( x\right) ,x\in \mathbb{R}^{3}\diagdown G_{1},  \label{4.32}
\end{equation}
\begin{equation}
\widehat{V}\left( x,t\right) \mid _{S_{\infty }}=\eta \left( x,t\right) .
\label{4.33}
\end{equation}
Recall that both functions $V_{j}\in H^{2}\left( \mathbb{R}^{3}\times \left(
0,T\right) \right) ,$ $\forall T>0.$ On the other hand, the standard energy
estimate tells us that the problem (\ref{4.31})-(\ref{4.33}) has at most one
solution $\widehat{V}\in H^{1}\left( \left( \mathbb{R}^{3}\diagdown
G_{1}\right) \times \left( 0,T\right) \right) ,$ $\forall T>0.$ Hence, $%
V_{1}\left( x,t\right) =V_{2}\left( x,t\right) =V\left( x,t\right) $ for $%
\left( x,t\right) \in \left( \mathbb{R}^{3}\diagdown G_{1}\right) \times
\left( 0,\infty \right) .$

Denote $\varphi \left( x,t\right) =\partial _{\nu }V\left( x,t\right) \mid
_{S_{\infty }}$, where $\nu $ is the unit normal vector at $S$ pointing
outside of the domain $G_{1}.$ For $j=1,2$ consider functions $W_{j}\left(
x,t\right) =\partial _{t}V_{j}\left( x,t\right) .$ Then 
\begin{equation}
\partial _{t}^{2}W_{j}=\Delta W_{j}-q_{j}\left( x\right) W_{j},\left(
x,t\right) \in G_{1}\times \left( 0,\infty \right) ,  \label{4.23}
\end{equation}
\begin{equation}
W_{j}\left( x,0\right) =g\left( x\right) ,\partial _{t}W_{j}\left(
x,0\right) =0,  \label{4.24}
\end{equation}
\begin{equation}
W_{j}\left( x,t\right) \mid _{S_{\infty }}=\eta _{t}\left( x,t\right) ,
\label{4.25}
\end{equation}
\begin{equation}
\partial _{\nu }W_{j}\left( x,t\right) \mid _{S_{\infty }}=\varphi
_{t}\left( x,t\right) .  \label{4.251}
\end{equation}
In addition, (\ref{2.12}) implies that 
\begin{equation}
W_{j}\in C^{3}\left( \overline{\Omega }\times \left[ 0,T\right] \right)
,\forall T>0.  \label{4.26}
\end{equation}
Finally, using Theorem 4.7 of \cite{Klib5}, we obtain that relations (\ref%
{4.23})-(\ref{4.26}) imply that $q_{1}\left( x\right) =q_{2}\left( x\right) $
in $\Omega .$ To finish the proof, we refer to (\ref{4.211}). $\square $

\textbf{Remark 2}. Let the number $T>diam\left( G_{1}\right) /2.$ Then one
can replace in (\ref{4.23}) the time interval $\left( 0,\infty \right) $
with $\left( 0,T\right) $ and also replace in (\ref{4.25}), (\ref{4.251}) $%
S_{\infty }$ with $S_{T}=S\times \left( 0,T\right) .$ These new conditions (%
\ref{4.23}), (\ref{4.25}) and (\ref{4.251}) together with (\ref{4.24}) and (%
\ref{4.26}) still imply that $q_{1}\left( x\right) =q_{2}\left( x\right) $
in $\Omega ,$ see either Theorem 1.10.5.2 of \cite{BK}, or Theorem 3.2.2 of 
\cite{KT}, or Theorem 3.2 of \cite{Kl16}.

\subsection{Proof of Theorem 4}

\label{sec.5.3}

Consider again an arbitrary point $\widetilde{x}\in S.$ For $j=1,2$ let $%
h_{j}\left( k\right) =v_{s,j}\left( \widetilde{x},k\right) $ and $\widehat{h}%
_{j}\left( t\right) =\mathcal{F}^{-1}\left( h_{j}\right) .$ Using (\ref%
{2.100}) and (\ref{2.150}), we obtain similarly with (\ref{4.181}) 
\begin{equation}
\widehat{h}_{1}\left( t\right) =\widehat{h}_{2}\left( t\right) =V_{s}\left( 
\widetilde{x},t\right) ,\text{ }\forall t\in \left( 0,\varepsilon \right) .
\label{4.27}
\end{equation}
We again apply arguments, which are completely analogous to those of section
3.3. Also, we keep the same notations (\ref{4.3}), (\ref{4.4}) and (\ref%
{4.10}). Hence, we obtain from (\ref{4.27}) the following analog of (\ref%
{4.190}) 
\begin{equation}
\int\limits_{0}^{t}V_{s}\left( \widetilde{x},t-\tau \right) \lambda \left(
\tau \right) d\tau =0,\text{ }\forall t\in \left( 0,\varepsilon \right) .
\label{4.28}
\end{equation}
Differentiate (\ref{4.28}) four times and use (\ref{2.12}) and (\ref{2.154}%
). We obtain the following analog of\ the Volterra integral equation (\ref%
{4.191}) 
\begin{equation*}
\lambda \left( t\right) -\frac{1}{\left( qg\right) \left( \widetilde{x}%
\right) }\int\limits_{0}^{t}\partial _{t}^{4}V_{s}\left( \widetilde{x}%
,t-\tau \right) \lambda \left( \tau \right) d\tau =0,\text{ }\forall t\in
\left( 0,\varepsilon \right) .
\end{equation*}
Hence, similarly with subsection 3.3 we conclude that $\lambda \left(
t\right) =0,$ $\forall t\geq 0,$ which implies that $v_{s,1}\left( 
\widetilde{x},k\right) =v_{s,2}\left( \widetilde{x},k\right) ,\forall k\in 
\mathbb{R}.$ Since $\widetilde{x}\in S$ is an arbitrary point, then 
\begin{equation}
v_{s,1}\left( x,k\right) =v_{s,2}\left( x,k\right) ,\text{ }\forall x\in
S,\forall k\in \mathbb{R}.  \label{4.29}
\end{equation}
Since $v_{j}\left( x,k\right) =v_{s,j}\left( x,k\right) +v_{0}\left(
x,k\right) ,$ then (\ref{4.29}) implies that (\ref{4.21}) is valid again.
The rest of the proof is the same as the proof of Theorem 3 after (\ref{4.21}%
). $\square $

\begin{center}
\textbf{Acknowledgment}
\end{center}

This research was supported by US Army Research Laboratory and US Army
Research Office grant W911NF-11-1-0399.

\bigskip

Michael Victor Klibanov

University of North Carolina at Charlotte

Department of Mathematics and Statistics

Charlotte, NC 28223

U.S.A.

mklibanv@uncc.edu

\end{document}